\documentclass[aps,prc,twocolumn,superscriptaddress,floatfix]{revtex4-1}
\usepackage{graphicx}
\usepackage{hyperref}
\usepackage{color}
\usepackage{changes}
\usepackage[T1]{fontenc}
\usepackage{amsmath,mathtools}

\begin{document}

\title{Challenges in QCD matter physics  --\\The scientific programme of the Compressed Baryonic Matter experiment at FAIR}




\author{T.~Ablyazimov}                 
\affiliation{Laboratory of Information Technologies, Joint Institute for Nuclear Research (JINR-LIT), Dubna, Russia}
\author{A.~Abuhoza}                    
\affiliation{GSI Helmholtzzentrum f\"{u}r Schwerionenforschung GmbH (GSI), Darmstadt, Germany}
\affiliation{also: King Abdulaziz City for Science and Technology (KACST), Riyadh, Saudi Arabia}
\author{R.P.~Adak}                     
\affiliation{Department of Physics, Bose Institute, Kolkata, India}
\author{M.~Adamczyk}                   
\affiliation{Marian Smoluchowski Institute of Physics, Jagiellonian University, Krak\'{o}w, Poland}
\author{K.~Agarwal}                    
\affiliation{Physikalisches Institut, Eberhard Karls Universit\"{a}t T\"{u}bingen, T\"{u}bingen, Germany}
\author{M.M.~Aggarwal}                 
\affiliation{Department of Physics, Panjab University, Chandigarh, India}
\author{Z.~Ahammed}                    
\affiliation{Variable Energy Cyclotron Centre (VECC), Kolkata, India}
\author{F.~Ahmad}                      
\affiliation{Department of Physics, University of Kashmir, Srinagar, India}
\author{N.~Ahmad}                      
\affiliation{Department of Physics, Aligarh Muslim University, Aligarh, India}
\author{S.~Ahmad}                      
\affiliation{Department of Physics, University of Kashmir, Srinagar, India}
% ...

\collaboration{CBM Collaboration}
\noaffiliation

\date{\today}

\begin{abstract}
Substantial experimental and theoretical efforts worldwide are devoted to explore the phase diagram of strongly interacting matter. 
At LHC and top RHIC energies, QCD matter is studied at very high temperatures and nearly vanishing net-baryon densities. 
There is evidence that a Quark-Gluon-Plasma (QGP) was created at experiments at RHIC and LHC. 
The transition from the QGP back to the hadron gas is found to be a smooth cross over. 
For larger net-baryon densities and lower temperatures, it is expected that the QCD phase diagram exhibits a rich structure, such as a first-order phase transition between hadronic and partonic matter which terminates in a critical point, or exotic phases like quarkyonic matter. 
The discovery of these landmarks would be a breakthrough in our understanding of the strong interaction and is therefore in the focus of various high-energy heavy-ion research programs. 
The Compressed Baryonic Matter (CBM) experiment at FAIR will play a unique role in the exploration of the QCD phase diagram in the region of high net-baryon densities, because it is designed to run at unprecedented interaction rates. 
High-rate operation is the key prerequisite for high-precision measurements of multi-differential observables and of rare diagnostic probes which are sensitive to the dense phase of the nuclear fireball. 
The goal of the CBM experiment at SIS100 ($\sqrt{s_{NN}} = 2.7- 4.9$~GeV) is to discover fundamental properties of QCD matter: 
the phase structure at large baryon-chemical potentials ($\mu_B >500$~MeV), effects of chiral symmetry, and the equation-of-state at high density as it is expected to occur in the core of neutron stars.
In this article, we review the motivation for and the physics programme of CBM, including activities before the start of data taking in 2024, in the context of the worldwide efforts to explore high-density QCD matter.

\end{abstract}

\maketitle

\section{Probing QCD Matter with Heavy-Ion Collisions}
\label{sec:intro}

\begin{figure}[htbp]
\begin{center}
\includegraphics[width=1.0\linewidth]{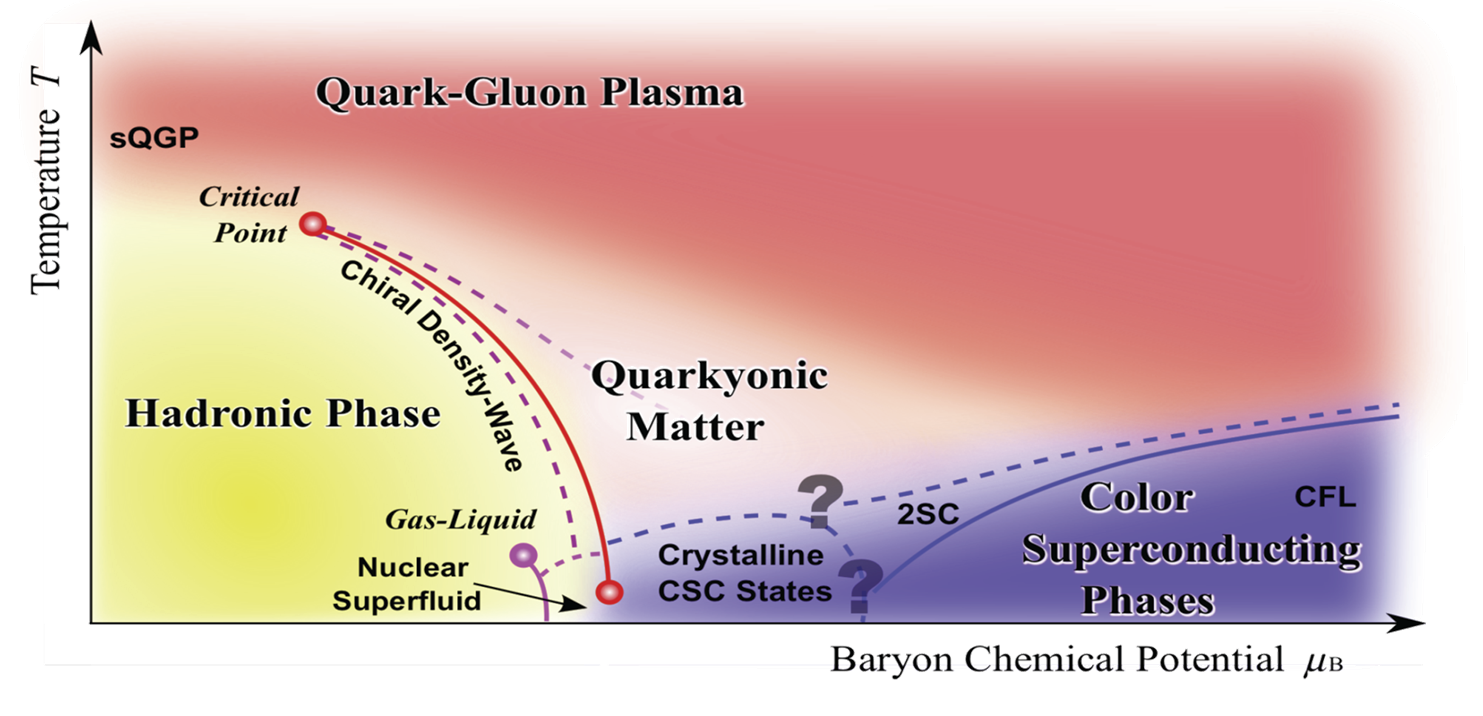}
\caption{Sketch of the phase diagram for strongly interacting matter (taken from~\cite{fukushima11}).}
\label{fig:phasediagram}
\end{center}
\end{figure}

Heavy-ion collision experiments at relativistic energies create extreme states of strongly interacting matter and enable their investigation in the laboratory. 
Figure~\ref{fig:phasediagram} illustrates the conjectured phases of strongly interacting matter and their boundaries in a diagram of temperature versus  baryon chemical potential~\cite{fukushima11}. 

Experiments at LHC and top RHIC energies explore the QCD phase diagram in the transition region between Quark-Gluon-Plasma (QGP) and hadron gas at small baryon chemical potentials, where matter is produced with almost equal numbers of particles and antiparticles. 
This region resembles the situation in the early universe. 
While cooling, the system hadronizes, and finally freezes out chemically at a temperature around 160 MeV~\cite{becattini13,stachel14}. 
This temperature coincides with the transition temperature predicted by first-principle Lattice QCD calculations~\cite{borsanyi10,basasov12}, which find a smooth crossover from partonic to hadronic matter~\cite{aoki06}. 
Lattice QCD calculations for finite baryon chemical potential are still suffering from the so-called sign problem, which makes the standard Monte-Carlo methods no longer applicable, and are not yet able to make firm predictions on possible phase transitions at large baryon chemical potentials. 
On the other hand, effective-model calculations predict structures in the QCD phase diagram at large baryon chemical potentials, like a critical endpoint followed by a first-order phase transition~\cite{kashiwa08,luecker13,tawfik15}. 
The development of a mixed phase of hadrons and quarks is e.g.\ predicted by a non-local \mbox{3-flavor} Nambu--Jona-Lasinio model calculation of a neutron star for densities around $5 \rho_0$, with a transition to pure quark matter above $8 \rho_0$. 
This calculation is able to reproduce a two-solar mass neutron star~\cite{orsaria14}. 
Moreover, a quarkyonic phase is predicted which has properties of both high density baryonic matter and deconfined and chirally symmetric quark matter~\cite{mclerran07,mclerran09}. 
Other scenarios discussed for matter at extreme densities include colour-flavour locking~\cite{alford99} and skyrmion matter~\cite{lee03}.

The experimental discovery of landmarks like 
a first-order phase transition or a critical point
in the QCD phase diagram would be a major breakthrough in our understanding of the strong interaction in the non-perturbative regime, with fundamental consequences for our knowledge on the structure of neutron star cores, chiral symmetry restoration, and the origin of hadron masses.

\begin{figure}[htbp]
\begin{center}
\includegraphics[width=1.0\linewidth]{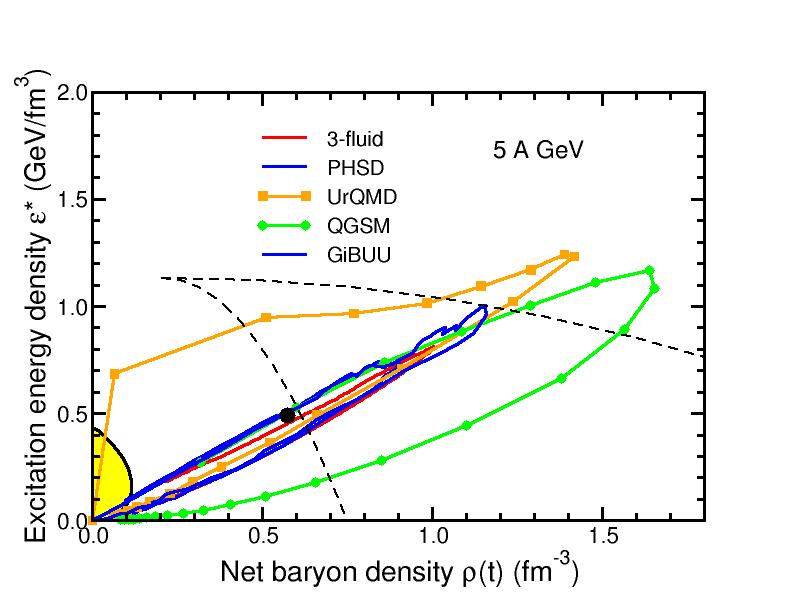}
\includegraphics[width=1.0\linewidth]{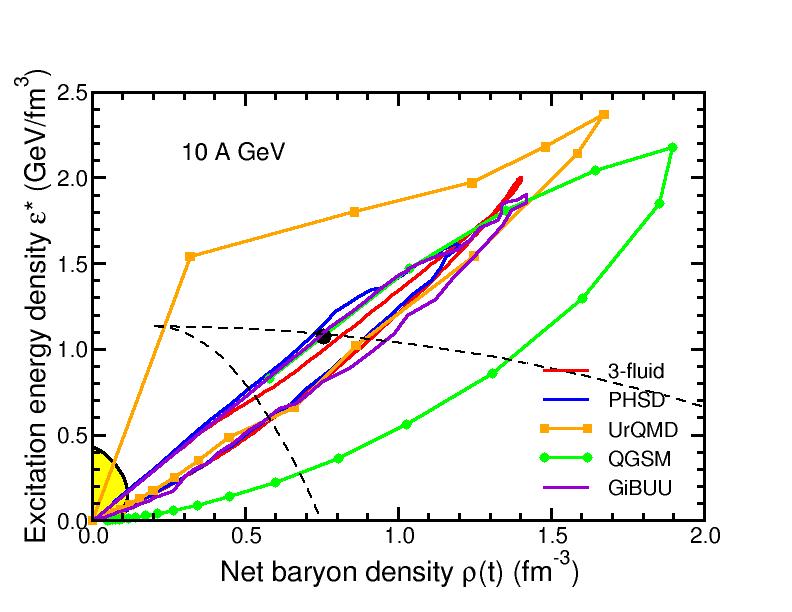}
\caption{
Time evolution of the excitation energy density $\epsilon^*$ versus the net-baryon density $\rho$ in the center of the fireball 
for central Au+Au collisions at beam energies of 5$A$~GeV (upper panel) and 10$A$~GeV (lower panel) calculated by various transport codes and a hydrodynamic model~\cite{arsene07,friman11}. 
The excitation energy density is defined as $ \epsilon^* = \epsilon - m_N \rho$ (see text). 
The full symbols on the curves for UrQMD and QGSM indicate time steps of 1~fm/$c$.
The dashed lines enclose the regions of phase coexistence~\cite{toneev03}.
The yellow zone denotes post-freezeout streaming.
}
\label{fig:trajectories}
\end{center}
\end{figure}

Heavy-ion collisions at moderate beam energies are well suited to provide high net-baryon densities. 
This is illustrated in Fig.~\ref{fig:trajectories}, where the excitation energy density in the center of the collision zone is shown as a function of the net-baryon density for central Au+Au collisions at beam energies of 5$A$ and 10$A$~GeV as predicted by several transport models and a hydrodynamic calculation~\cite{arsene07,friman11}. 
The excitation energy is defined as 
\mbox{$ \epsilon^*(t) = \epsilon(t) - m_N \rho(t)$}
with $\epsilon(t)$ the energy density and $m_N \rho(t)$ the mass density. 
The solid lines correspond to the time evolution of the system; they turn in a clockwise sense, and the dots on the curves labelled UrQMD and QGSM correspond to steps of 1 fm/$c$ in collision time. 
The dashed lines enclose the expected region of phase coexistence~\cite{toneev03}.

According to these model calculations, the density in the center of the fireball exceeds 6 times saturation density $\rho_0$ at a beam energy of 5$A$~GeV, and at 10$A$~GeV even a density above $8 \rho_0$ is reached. 
At such densities, the nucleons are expected to fuse and form large quark bags. 
The calculations predict that the dense fireball spends a comparatively long time within the phase coexistence region at energies around 5$A$~GeV and goes beyond this region with increasing beam energy. 

High-density matter as produced in nuclear collisions at FAIR energies also opens the possibility to search for multi-strange hypernuclei.  
Experimental data on such objects are very scarce; detailed studies of their production will give information on the hyperon-hyperon interaction which is essential for the understanding of cores of neutron stars. 
Models predict the FAIR energy range to be particularly well suited for such studies. 
This also holds for the search for exotic composite objects carrying multiple units of strangeness like kaonic clusters or multi-strange di-baryons, the existence of which is still an open issue in high-energy physics.

In conclusion, the systematic and comprehensive exploration of the QCD phase diagram in the region of high-net baryon densities using heavy-ion collisions at SIS100 beam energies (up to 11$A$~GeV for Au ions)  and measuring diagnostic probes never observed before in this energy regime will have a large discovery potential. 
In particular, the CBM experiment operated at intermediate beam energies will be able to address the following fundamental questions: 
\begin{itemize}
\item{
What is the equation of state of QCD matter at high net-baryon densities, and what are the relevant degrees of freedom at these densities? 
Is there a phase transition from hadronic to quark-gluon matter, or a region of phase coexistence? 
Do exotic QCD phases exist?
}
\item{
To what extent are the properties of hadrons modified in dense baryonic matter? 
Are we able to find indications of chiral symmetry restoration?
}
\item{
How far can we extend the chart of nuclei towards the third (strange) dimension by producing single and double strange hypernuclei? 
Does strange matter exist in the form of heavy multi-strange objects?
}
\end{itemize}

The focus of the CBM experiment at FAIR is to study observables related to the physics cases mentioned above.
The equation-of-state can be studied by measuring (i) the collected flow of identified particles, which is generated by the density gradient of the early fireball, 
and (ii) by multi-strange hyperons, which are preferentially produced in the dense phase of the fireball via sequential collisions.
A phase transition from hadronic to partonic matter is expected to cause the following effects:
(i) multi-strange hyperons are driven into equilibrium at the phase boundary;
(ii) in case of a first-order phase transition, the excitation function of the fireball temperature -- measured by the invariant-mass spectra of lepton pairs -- should reflect a caloric curve.
A possible critical point should produce event-by-event fluctuations of conserved quantities such as strangeness, charge, and baryon number.
Modifications of hadron properties in dense baryonic matter and the onset of chiral symmetry restoration affect the invariant-mass spectra of di-leptons.
The measurement of (double-$\Lambda$) hyper-nuclei will provide information on the hyperon-nucleon and hyperon-hyperon interaction which will shed light on the hyperon puzzle in neutron stars.

A more detailed discussion of the relation between the various physics cases and observables is presented in section~\ref{sec:probes} of this article, together with a review of the current data situation and the discovery potential of the CBM experiment.
Before, we give a brief overview of the experimental landscape in section~\ref{sec:experiments} and of the CBM detector in section~\ref{sec:cbm}.
A general introduction into theoretical concepts and experimental programmes devoted to the exploration of the QCD phase diagram at high net-baryon densities can be found in the CBM Physics Book~\cite{friman11}.

\section{Experiments exploring high net-baryon densities}
\label{sec:experiments}

Most of the experimental observables which are sensitive to the properties of dense nuclear matter, like the flow of identified (anti-) particles, higher moments of event-by-event multiplicity distributions of conserved quantities, multi-strange (anti-) hyperons, di-leptons, and particles containing charm quarks are extremely statistics-demanding. 
Therefore, the key feature of successful experiments will be rate capability in order to measure these observables with high precision. 
The experimental challenge is to combine a large-acceptance fast detector and a high-speed data read-out system with high-luminosity beams.     

The QCD phase diagram at large baryon chemical potentials has been explored by pioneering heavy-ion experiments performed at AGS in Brookhaven and at low CERN-SPS beam energies. 
Because of the then available detector technologies these measurements were restricted to abundantly produced hadrons and to di-electron spectra with strongly limited statistics. 
At the CERN-SPS, the NA61/SHINE experiment continues to search for the first-order phase transition by measuring hadrons using light and medium heavy ion beams~\cite{laszlo07}. 
This detector setup is limited to reaction rates of about 80~Hz. 
The existing HADES detector at SIS18 measures hadrons and electron pairs in heavy-ion collision systems with reaction rates up to 20~kHz. 
The STAR collaboration at RHIC has performed a beam energy scan from top energies down to $\sqrt{s_{NN}} = 7.7$~GeV, and plans to improve the statistical significance of the data in a second beam energy scan~\cite{starbes}.  
At beam energies above $\sqrt{s_{NN}} = 20$~GeV, the reaction rates of STAR are limited to about 800~Hz by the TPC read-out, and drop down to a few Hz at beam energies below $\sqrt{s_{NN}} = 8$~GeV because of the decreasing beam luminosity provided by the RHIC accelerator. 
At the Joint Institute for Nuclear Research (JINR) in Dubna, the fixed-target experiment BM@N is being developed at the Nuclotron to study heavy-ion collisions at gold beam energies up to about 4$A$~GeV. 
Moreover, at JINR the Nuclotron-based Ion Collider fAcility NICA with the Multi-Purpose Detector (MPD) is under construction~\cite{nica}. 
The NICA collider is designed to run at a maximum luminosity of $L = 10^{27} \mathrm{cm^{-2}s^{-1}}$ at collision energies between $\sqrt{s_{NN}} = 8$ and 11 GeV corresponding to a reaction rate of 6~kHz for minimum bias Au+Au collisions. 
The interaction rate at NICA decreases to about 100~Hz because of low luminosity at $\sqrt{s_{NN}} = 5$~GeV. 

\begin{figure}[htbp]
\begin{center}
\includegraphics[width=1.0\linewidth]{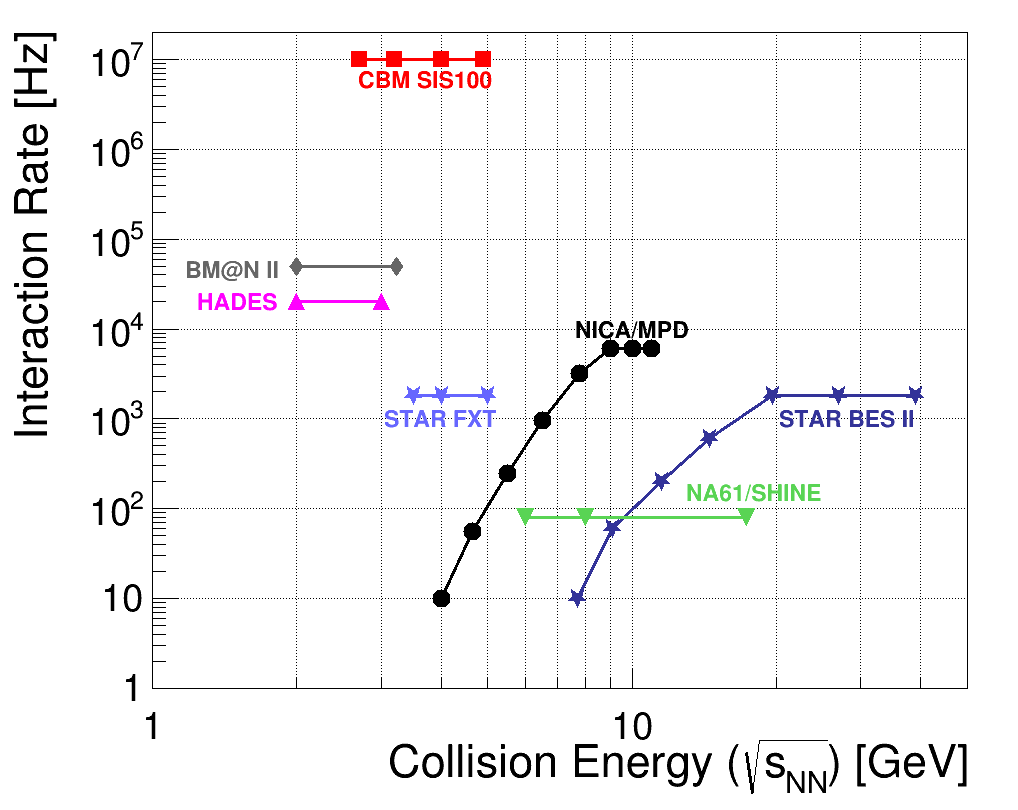}
\caption{Interaction rates achieved by existing and planned heavy-ion experiments as a function of center-of-mass energy~\cite{nica,montag14,michel11,odyniec13}. ``STAR FXT'' denotes the fixed-target operation of STAR.
High-rate experiments are also proposed at JPARC~\cite{sako15} and at SPS~\cite{dainese16}, but these are still in a conceptual stage.}
\label{fig:experiments}
\end{center}
\end{figure}

The Facility for Antiproton and Ion Research (FAIR), currently under construction in Darmstadt, will offer the opportunity to study nuclear collisions at extreme interactions rates. 
The FAIR Modularized Start Version (MSV) comprises the SIS100 ring which provides energies for gold beams up to 11$A$~GeV ($\sqrt{s_{NN}} = 4.9$~GeV), for Z=N nuclei up to 15$A$~GeV, and for protons up to 30~GeV. 
In order to reach higher energies, a booster ring is needed. 
The space for this second accelerator is already foreseen in the ring tunnel building. 
The rate capabilities of existing and planned heavy-ion experiments are presented in Fig.~\ref{fig:experiments} as a function of center-of-mass energy.

The research program on dense QCD matter at FAIR will be performed by the experiments CBM and HADES. 
The HADES detector, with its large polar angle acceptance ranging from 18 to 85 degrees~\cite{agakishiev09}, is well suited for reference measurements with proton beams and heavy ion collision systems with moderate particle multiplicities, i.e.\ Ni+Ni or  Ag+Ag collisions at the lowest SIS100 energies. 
Electron pairs and hadrons including multi-strange hyperons can be reconstructed with HADES. 

The CBM detector~\cite{friman11} is a fixed target experiment designed to run at extremely high interaction rates up to 
10~MHz for selected observables such as J/$\psi$, at 1-5~MHz for multi-strange hyperons and dileptons, and at 100~kHz without any online event selection. 
The CBM detector system will accept polar emission angles between 2.5 and 25 degrees in order to cover mid-rapidity and the forward rapidity hemisphere for symmetric collision systems over the FAIR energy range. 
The combination of high-intensity beams with a high-rate detector system and sufficient beam time provides worldwide unique conditions for a comprehensive study of QCD matter at the highest net-baryon densities achievable in the laboratory.

\section{The CBM experiment at  FAIR}
\label{sec:cbm}

As discussed above, the SIS100 energy range is well suited to produce and to investigate strongly interacting matter at densities as they are expected to exist in the core of neutron stars. 
This opens the perspective to study the fundamental questions raised above with a dedicated experiment which is ideally suited to measure rare diagnostic probes of dense matter with high accuracy. 
In the following we discuss the detector requirements and highlights of the physics program. 
   
The CBM detector has been designed as a multi-purpose device which will be capable to measure hadrons, electrons and muons in elementary nucleon and heavy-ion collisions over the full FAIR beam energy range. 
Therefore, no major adjustments have to be made to optimize the experiment for SIS100 beams. 
A staging scenario is, however, foreseen for some detector systems and for the DAQ system. 

In order to perform high-precision multi-differential measurements of rare probes the experiment should run at event rates of 100~kHz up to 10~MHz for several months per year. 
To filter out weakly decaying particles like hyperons or D mesons, no simple trigger signal can be generated. 
Instead, the full events have to be reconstructed, and the decay topology has to be identified online by fast algorithms running on a high-performance computing farm hosted by the GSI GreenIT cube. 
To utilize maximum rates, the data acquisition is based on self-triggered front-end electronics. 

\begin{figure*}[htbp]
\begin{center}
\includegraphics[width=0.8\linewidth]{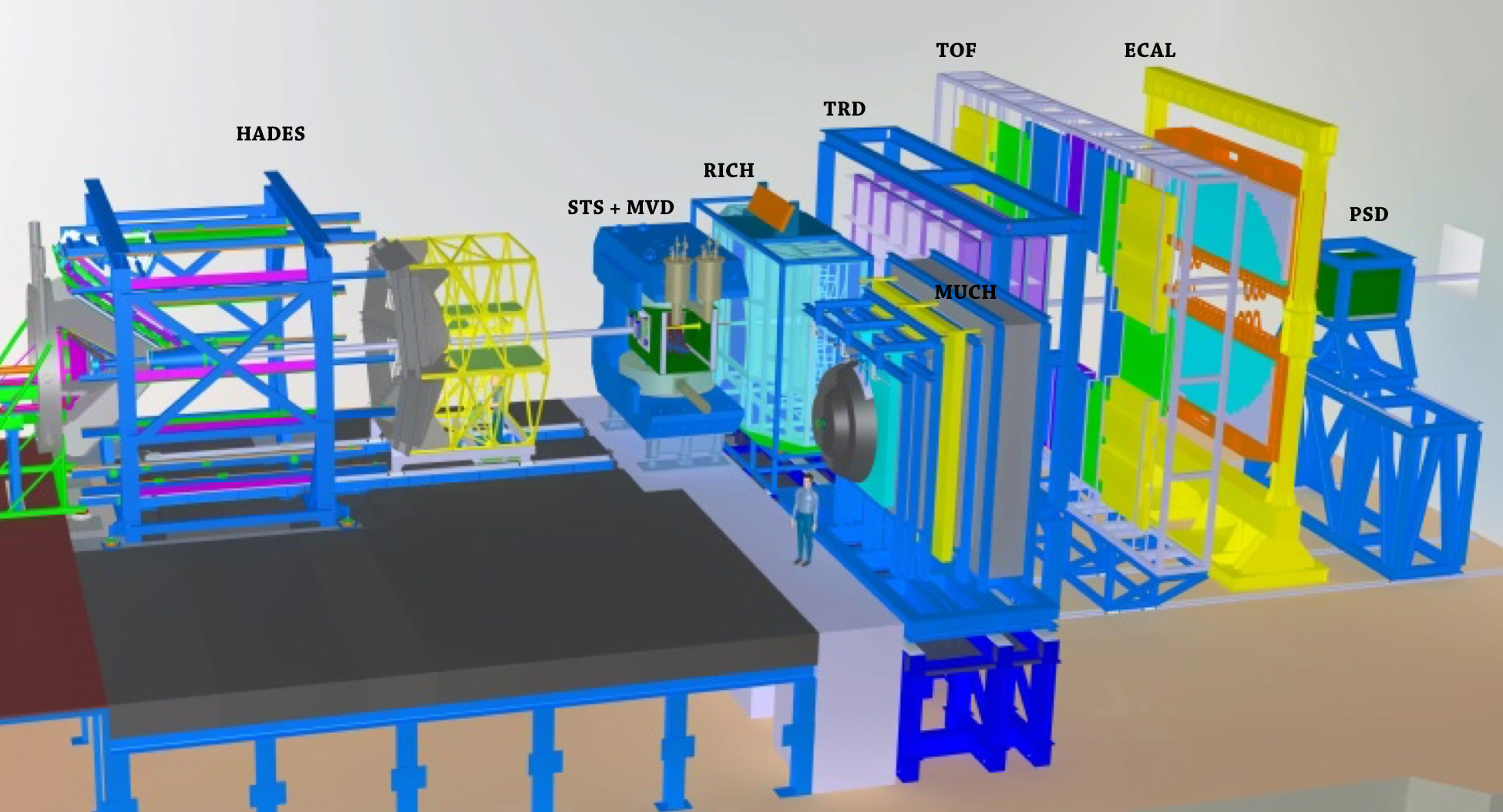}
\caption{
The CBM experimental setup together with the HADES detector (left). 
Each setup has its own target. 
During HADES operation, the beam will be stopped by a beam dump in front of CBM.  
The CBM components are described in the text. 
For muon measurements, the RICH will be exchanged by the MuCh which is shown in a parking position to the right of the beam axis. 
}
\label{fig:cbmsetup}
\end{center}
\end{figure*}

The CBM experimental setup is depicted in Fig.~\ref{fig:cbmsetup} and comprises the following components:
\begin{itemize}
\item a superconducting dipole magnet,
\item a Micro Vertex Detector (MVD) consisting of four layers of silicon monolithic active pixel sensors,
\item a Silicon Tracking System (STS) based on double-sided silicon micro-strip sensors arranged in eight stations inside 
a dipole magnet,
\item a Time-of-Flight wall (TOF) based on Multi-Gap Resistive Plate Chambers (MRPC) with low-resistivity glass,
\item a Ring Imaging Cherenkov (RICH) detector comprising a $\mathrm{CO_2}$ radiator and a UV photon detector realized with multi-anode photomultipliers for electron identification,
\item a Transition Radiation Detector (TRD) for pion suppression, particle tracking, and identification using specific energy loss, 
\item a Muon Chamber (MuCh) system for muon identification consisting of a set of gaseous micro-pattern chambers sandwiched between hadron absorber plates made of graphite and iron,
\item an Electromagnetic Calorimeter (ECAL) for the measurement of photons,
\item a Projectile Spectator Detector (PSD) for event characterization,
\item a First-Level-Event-Selection (FLES) system for online event reconstruction and selection.
\end{itemize}

The preparation of the experiment is well advanced. The Technical Design Reports (TDRs) of the Dipole Magnet, the STS, the TOF wall, the RICH, the MuCh and the PSD have been approved~\cite{magnetTdr,stsTdr,tofTdr,richTdr,muchTdr,psdTdr}, and the TDRs of the MVD, the TRD and the FLES are in progress. According to the schedule, the CBM experiment will be ready to take the first beams from SIS100 in 2024.

\section{Probes of high-density QCD matter}
\label{sec:probes}

The theoretical understanding of the properties of strongly interacting matter at large net-baryon densities is still poor. 
The scientific progress in this field is mainly driven by new experimental results. 
Owing to the complexity of the final state of heavy-ion reactions, the extraction of significant information requires systematic measurements like excitation functions, system size dependencies and multi-differential phase-space distributions of identified particles, including flow, event-by-event fluctuations, and other types of correlations. 
This task is even more challenging for high-statistics measurements of rare and penetrating probes. 
In the following we discuss the most promising observables in some detail. 

\subsection{Collectivity}

The collective flow of hadrons is driven by the pressure gradient created in the early fireball and provides information on the dense phase of the collision 
(for an overview, see~\cite{herrmann99,oeschler10} and references therein). 
Flow effects can be characterized by the azimuthal distribution of the emitted particles
$dN/d\phi = C \left( 1 + v_1 \cos(\phi) + v_2 \cos(2 \phi) + ... \right)$,
where $\phi$ is the azimuthal angle relative to the reaction plane, and the coefficients $v_1$ and $v_2$ represent the strengths of the directed (in-plane) and the elliptic flow, respectively.
At SIS100 energies, the proton flow has been measured between 2$A$ and 10.7$A$~GeV in Au+Au collisions~\cite{pinkenburg99}. 
These data have been compared to the results of transport model calculations in order to extract information on the nuclear matter equation of state (EOS)~\cite{danielewicz02}. 
Moreover, a large flow of kaons has been observed in Au+Au collisions at 6$A$~GeV~\cite{chung00}.
At SIS18 (1$A$ - 2$A$~GeV), exploratory measurements of kaon flow have been performed by the FOPI and KaoS experiments~\cite{shin98,zinhuk14}.

\begin{figure}[htbp]
\begin{center}
\includegraphics[width=1.0\linewidth]{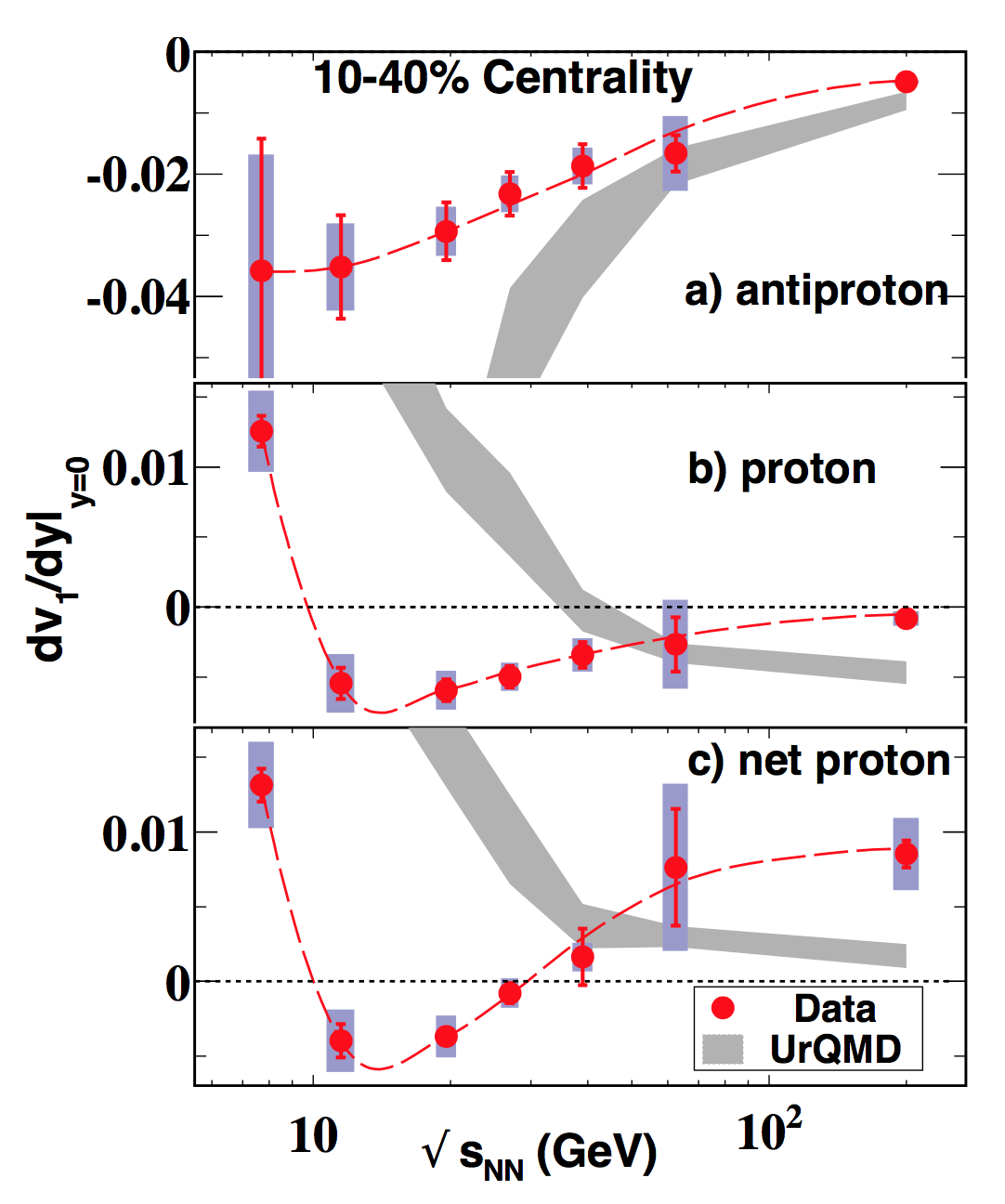}
\caption{Directed flow slope $\mathrm{d}v_1/\mathrm{d}y$ near mid-rapidity versus beam energy for intermediate-centrality Au+Au. Panels (a), (b) and (c) depict measured antiprotons, protons, and net protons, respectively, along with UrQMD 
calculations (grey bands)~\cite{adamczyk14}.}
\label{fig:starv1}
\end{center}
\end{figure}

Recently, the STAR collaboration has measured the directed flow for protons and antiprotons~\cite{adamczyk14} and the elliptic flow for particles and antiparticles~\cite{starbes} in Au+Au collisions at energies from 
$\sqrt{s_{NN}} = 62.4$~GeV down to $\sqrt{s_{NN}} = 7.7$~GeV.
Figure~\ref{fig:starv1} shows the measured slope of the directed flow of antiprotons, protons and net-protons together with the results of an UrQMD calculation. 
The directed flow is sensitive to the details of the phase transition, the softening of the QCD matter EOS, and is an important observable for clarifying the role of partonic degrees of freedom~\cite{steinheimer14}. 
Transport models such as UrQMD are challenged to reproduce details of the energy dependence and magnitude of the $v_1$ slope measured by STAR.   

\begin{figure}[htbp]
\begin{center}
\includegraphics[width=1.0\linewidth]{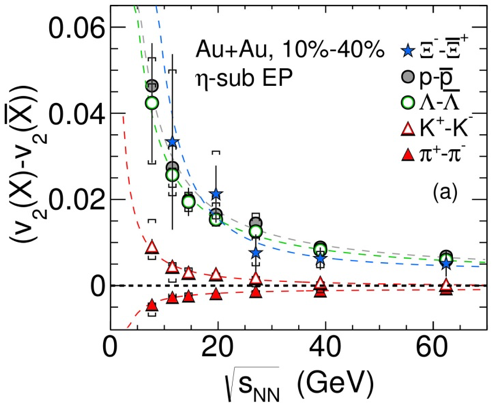}
\caption{
The difference in elliptic flow between particles and their corresponding antiparticles (see legend) as a function of 
$\sqrt{s_{NN}}$ for 10\% - 40\% central  Au+Au collisions as measured by the STAR collaboration at RHIC~\cite{starbes}. 
The systematic errors are indicated by the brackets. 
The dashed lines in the plot are fits with a power-law. 
}
\label{fig:starv2}
\end{center}
\end{figure}

Figure~\ref{fig:starv2} depicts the measured difference in elliptic flow $v_2$ for particles and antiparticles as a function of center-of-mass energy. 
The difference increases with increasing particle mass towards lower collision energies. 
This $v_2$ splitting was attributed to effects of the mean-field potential in both the partonic and the hadronic phase~\cite{xu14}.
On the other hand, it was argued that the baryon chemical potential is the determining factor for the observed particle type dependent splitting in $v_2$~\cite{hatta16}. 

At the lowest collisions energy of the RHIC beam-energy scan, which is close to the SIS100 energy range, 
$v_2$ measurements are only available for pions, protons, antiprotons, charged kaons, and (with poor precision) 
$\Lambda$/$\overline{\Lambda}$.
The CBM experiment will therefore dramatically improve the data situation by measuring the flow of identified particles in the FAIR energy range, including multi-strange hyperons and di-leptons. 
Of particular interest is the flow of particles not significantly suffering from rescattering like $\Omega$ hyperons or $\phi$ mesons, for which no experimental data exist.
These measurements will significantly contribute to our understanding of the QCD matter equation-of-state at neutron star core densities.

\subsection{Event-by-event fluctuations}

Event-by-event fluctuations of conserved quantities such as baryon number, strangeness and electrical charge can be related to the thermodynamical susceptibilities and hence provide insight into the properties of matter created in high-energy nuclear collisions. 
Lattice QCD calculations suggest that higher moments of these distributions are more sensitive to the phase structure of the hot and dense matter created in such collisions.  
Non-Gaussian moments (cumulants) of these fluctuations are expected to be sensitive to the proximity of the critical point since they are proportional to powers of the correlation length, with increasing sensitivity for higher-order moments.

\begin{figure}[htbp]
\begin{center}
\includegraphics[width=1.0\linewidth]{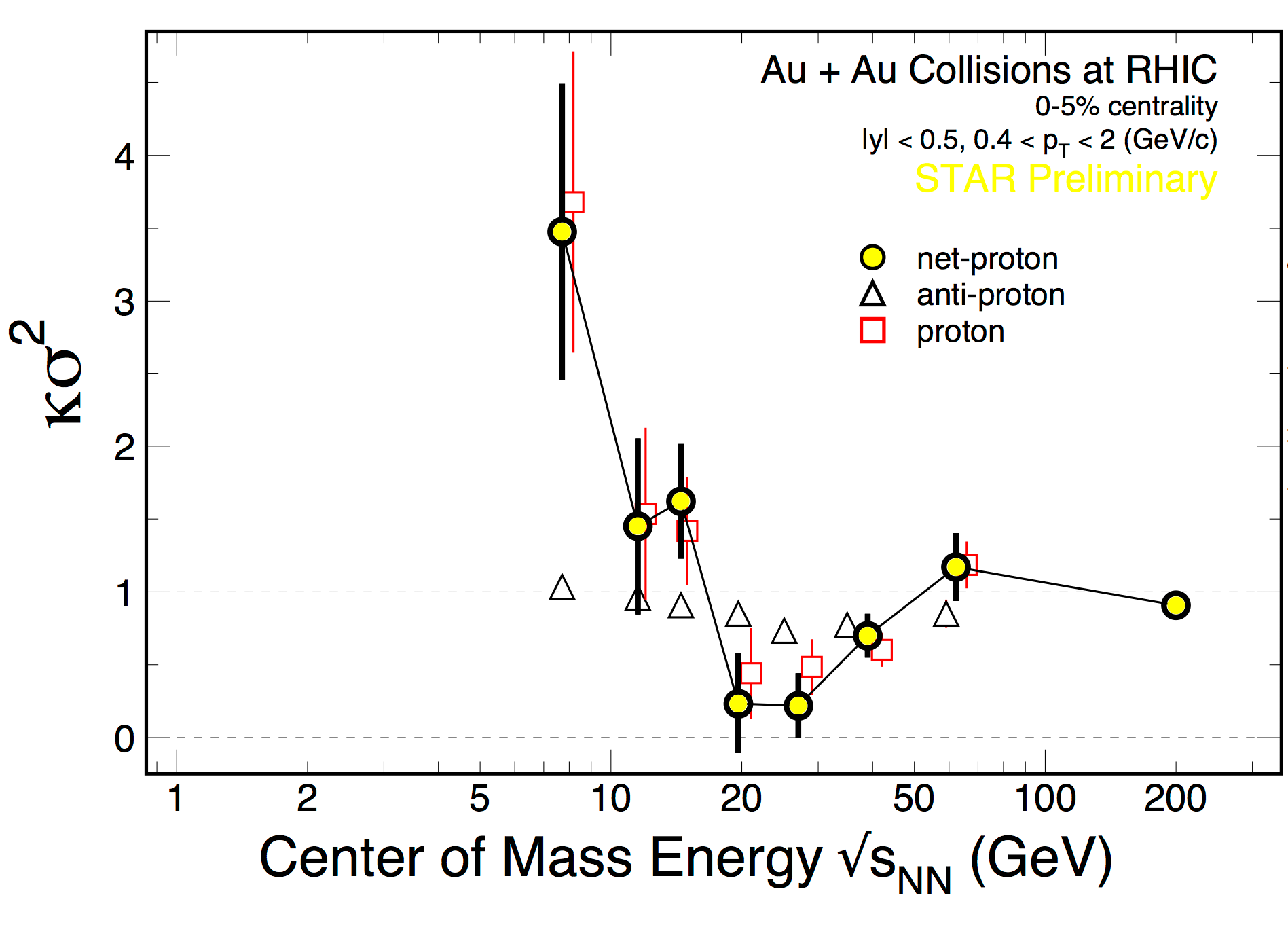}
\end{center}
\begin{center}
\caption{
Energy dependence of the product $\kappa \sigma^2$ (excess kurtosis times variance) of the net-proton multiplicity distribution (yellow circles) for top 0-5\% central Au+Au collisions. 
The Poisson expectation is denoted as dotted line at \mbox{$\kappa \sigma^2$ = 1}~\cite{luo14,thaeder16}. 
}
\label{fig:fluctuations}
\end{center}
\end{figure}

Measurements of event-by-event fluctuations have been performed by the NA49, PHENIX and STAR collaborations in order to search for the QCD critical point~\cite{alt09,anticic15-1,anticic15-2,mitchell15,adare16,adamczyk15}. 
Recent results from STAR are shown in Fig.~\ref{fig:fluctuations}, which depicts the volume-independent cumulant ratio $\kappa \sigma^2$ (excess kurtosis times squared standard deviation) of the net-proton multiplicity distribution as a function of the collision energy, measured in Au+Au collisions~\cite{luo14,thaeder16}. 
In the absence of a critical point, this quantity is found to be constant as a function of collision energy in various model calculations~\cite{karsch11,skokov13,garg13,luo10}.
The presence of a critical point is expected to lead to a non-monotonic behaviour of the $\kappa \sigma^2$ observable~\cite{stephanov11,chen15}.
For the most central collisions the STAR-BES data exhibit a deviation from unity at the lowest measured energy as expected for a critical behaviour. 
These results clearly call for a high-precision measurement of higher-order fluctuations at lower beam energies in order to search for the peak in $\kappa \sigma^2$.
   
Up to date no higher-order event-by-event fluctuations have been measured at SIS100 energies. 
The CBM experiment will, for the first time, perform a high-precision study of higher-order fluctuations at various beam energies in order to search for the elusive QCD critical point in the high net-baryon density region: 
\mbox{ $\sqrt{s_{NN}} = 2.7$ -- 4.9~GeV }
corresponding to \mbox{$\mu_B \simeq 800$ -- 500~MeV}. 

As recently pointed out, large clusters of nucleons might be important for the critical behaviour in the high net-baryon density region~\cite{bzdak16}. In addition, the density fluctuations arising from the criticality can also be accessed via the measurements of the yields of light nuclei such as deuterons, assuming coalescence to be the production mechanism. 
Precise measurements of the energy dependence of light nuclei production will further aid and complement to the critical point searches at the high baryon density region at FAIR.

\subsection{Strangeness}
Particles containing strange quarks are important probes of the excited medium created in heavy-ion collisions~\cite{koch86,gazdzicki99,tomasik16}.
At top SPS energy strange hadrons, including $\Omega$ and $\overline{\Omega}$, appear to be produced in chemical 
equilibrium~\cite{andronic10}. 
The equilibration of in particular $\Omega$ baryons could not be understood in terms of hadronic two-body relaxation processes in the limited life time of the fireball. 
It was thus taken as strong indication that the system had undergone a transition from a partonic phase to the hadronic final state, with the equilibration being driven by multi-body collisions in the high particle density regime near the phase boundary~\cite{pbm04}.

\begin{figure}[htbp]
\begin{center}
\includegraphics[width=1.0\linewidth]{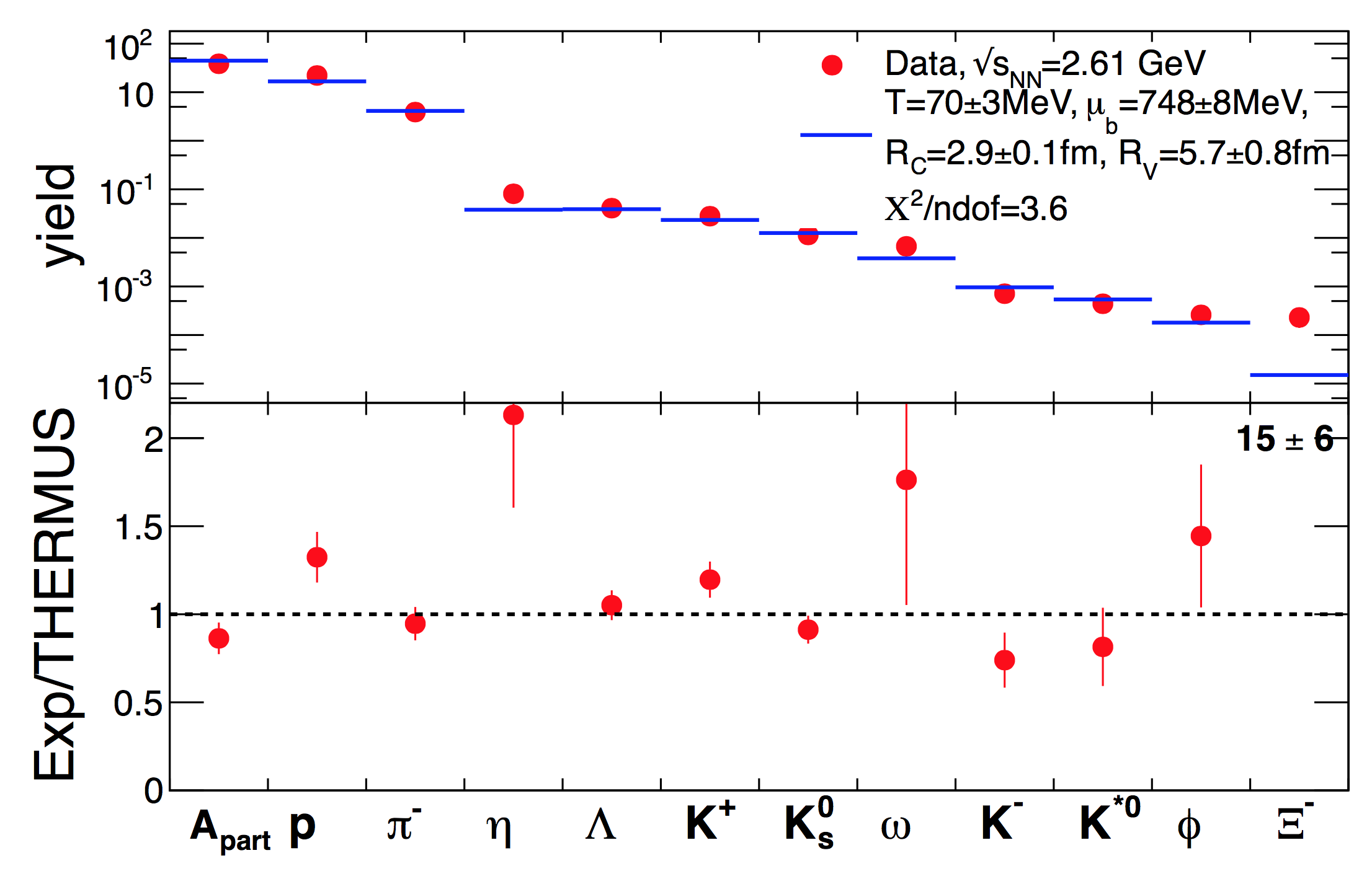}
\caption{
Yield of hadrons measured  in Ar + KCl collisions at an energy of 1.76$A$~GeV by the HADES 
collaboration (full symbols). 
The horizontal bars represent a fit by a thermo-statistical model~\cite{agakishiev09b}.
}
\label{fig:hadesyields}
\end{center}
\end{figure}

Agreement of the $\Omega$ baryon yield with thermal model calculations was found also at 40$A$~GeV in Pb+Pb collisions at the SPS~\cite{andronic09}, although the data statistics is rather poor. 
In the AGS (SIS100) energy range, only about 300 $\Xi^-$ hyperons have been measured in Au+Au collisions at 6$A$~GeV~\cite{chung03}. 
Figure~\ref{fig:hadesyields} depicts the yield of hadrons measured in Ar + KCl collisions at an energy of 1.76$A$~GeV together with the result of a statistical model calculation~\cite{agakishiev09b}. 
The measured yield of $\Xi^-$ hyperons exceeds the model prediction by about a factor of 20, indicating that $\Xi^-$ is far off chemical equilibrium. 
High-precision measurements of excitation functions of multi-strange hyperons in A+A collision with different mass numbers A at SIS100 energies will allow to study the degree of equilibration of the fireball, and, hence, open the possibility to find a signal for the onset of deconfinement in QCD matter at high net-baryon densities.   

According to hadronic transport models, which do not feature a partonic phase, multi-strange (anti-)hyperons are produced in sequential collisions involving kaons and lambdas, and, therefore, are sensitive to the density in the fireball. 
This sensitivity is largest at lower beam energies close to or even below the production threshold in elementary collisions, and is expected to shed light on the compressibility of nuclear matter.

The CBM experiment will open a new era of multi-differential precision measurements of strange hadrons including multi-strange (anti-) hyperons. 
The expected particle yields are sufficient to study with excellent statistical significance the production and propagation of heavy strange and anti-strange baryons up to $\Omega^+$ in dense nuclear matter. 
Also excited hyperon states can be identified. 
Moreover, it will be possible to study hyperon-nucleon and hyperon-hyperon correlations in order to explore the role of hyperons in neutron stars, which is of utmost importance with respect to the difficulty to reconcile the measured masses of neutron stars with the presence of hyperons in their interiors, the so-called hyperon puzzle~\cite{demorest10}.

\subsection{Lepton pairs}

Di-leptons emitted in collisions of heavy ions offer the unique opportunity to investigate the microscopic properties of 
strongly interacting matter~\cite{mclerran85,weldon90}. 
Virtual photons are radiated off during the whole time evolution of a heavy-ion collision. 
Once produced, they decouple from the collision zone and materialize as muon or electron pairs. 
Hence, leptonic decay channels offer the possibility to look into the fireball and to probe the hadronic currents of strongly interacting systems in a state of high temperature and density. 
For example, the low-mass continuum in the invariant mass spectrum of lepton pairs 
($M < 1 \, \mathrm{GeV}/c^2$) probes the in-medium $\rho$ spectral function as this meson saturates, according to vector meson dominance, the hadronic current in a hadron resonance gas~\cite{gale91}. 
Moreover, the excess yield of lepton pairs in this energy range is sensitive to both the temperature of the created matter and its lifetime (or more precisely its space-time extension). 
This observable is expected to be a measure of the fireball lifetime and to be sensitive to chiral symmetry restoration~\cite{hohler14}. 
The slope of the dilepton invariant mass distribution between 1 and 2.5~GeV/$c^2$ directly reflects the average temperature of the fireball~\cite{rapp16}. 
This measurement would also provide indications for the onset of deconfinement and the location of the critical endpoint. 
The flow of lepton pairs as function of their invariant mass would allow to disentangle radiation of the early partonic phase from the late 
hadronic phase~\cite{chatterjee07,deng11,mohanty12,gale15}. 
No di-lepton data have been measured in heavy-ion collisions at beam energies between 2$A$ and 40$A$~GeV.

\begin{figure}[htbp]
\begin{center}
\includegraphics[width=1.0\linewidth]{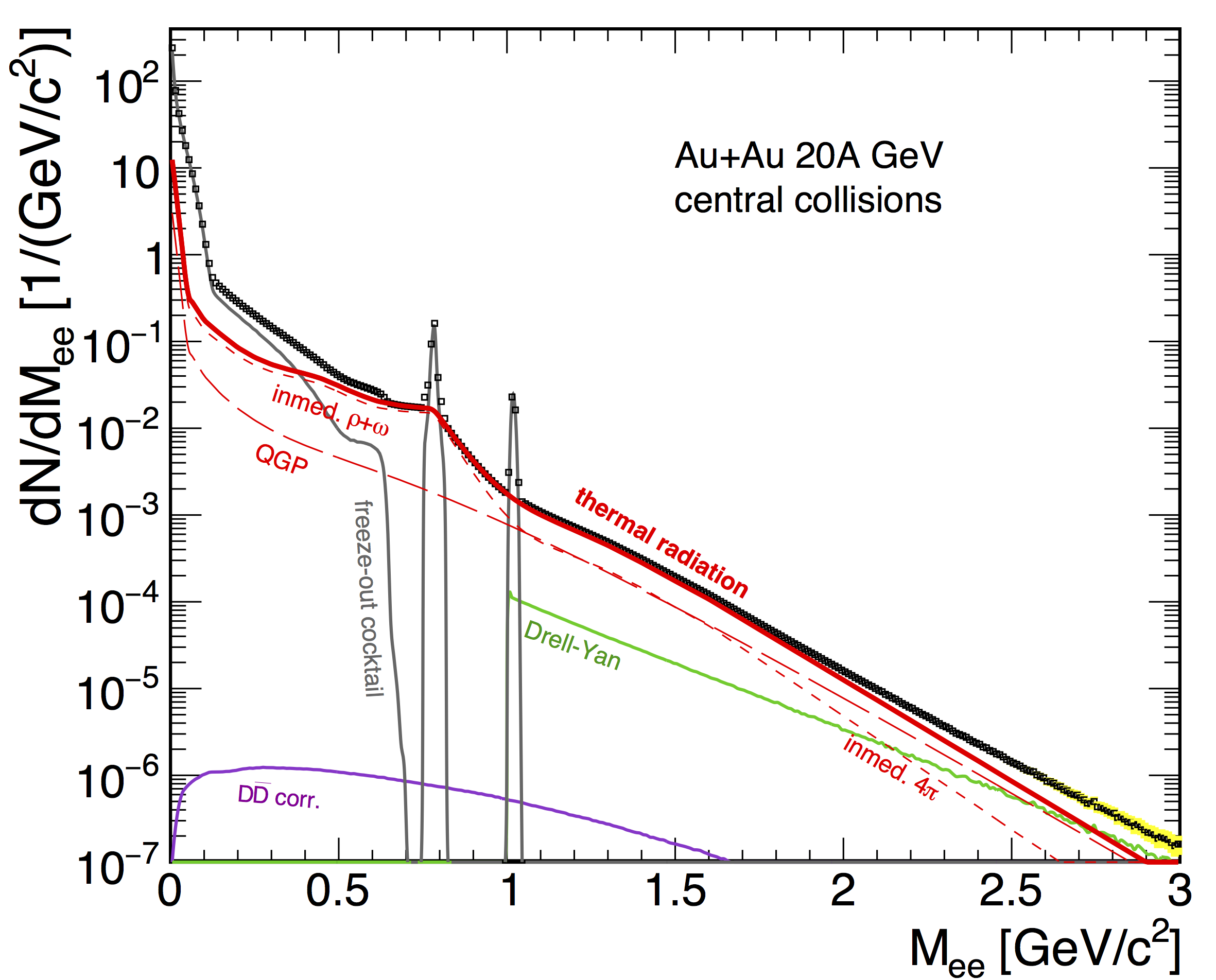}
\caption{
Invariant-mass spectrum of $e^+e^-$ pairs radiated from a central Au+Au collision at 20$A$ GeV. 
The solid red curve shows the contribution of the thermal radiation which includes in-medium $\rho$, $\omega$, 4-$\pi$ spectral functions and QGP spectrum calculated using the many-body approach of~\cite{rapp99}.
The freeze-out hadron cocktail (solid grey curve) is calculated using the Pluto event generator~\cite{froehlich07} and includes two-body and Dalitz decays of $\pi^0$, $\eta$, $\omega$, and $\phi$. 
Contributions of Drell-Yan (green solid curve) and correlated open charm (solid violet curve) have been simulated based on~\cite{bhaduri14}.
}
\label{fig:dielectrons}
\end{center}
\end{figure}

The CBM experiment will perform pioneering multi-differential measurements of lepton pairs over the whole range of invariant masses emitted from a hot and dense fireball. 
According to model calculations, various processes will contribute to the measured yield as shown in Fig.~\ref{fig:dielectrons}~\cite{rapp99}. 
The thermal radiation includes a broadened in-medium $\rho$ meson~\cite{friman93,chanfray93,leupold98,chanfray96,oset02}, 
radiation from the QGP~\cite{ding16}, and dileptons from multi-pion 
annihilation~\cite{hohler14,rapp16}. 
The latter reflects $\rho$-$a_1$ chiral mixing and therefore provides a direct link to chiral symmetry restoration.
The experimental challenges are the very low signal cross sections, decay probabilities of the order of $10^{-4}$, and the high combinatorial background. 
According to simulations, it will be possible to identify di-leptons in the relevant invariant mass regions with a signal-to-background ratio of at least S/B = 1/100. 
In this case one needs about 10000 signal pairs in order to determine the yield with a statistical accuracy of 10\%.  
The expected signal yield in 10 weeks of running the CBM experiment is higher by a factor of 100 -- 1000, depending on beam energy.

\begin{figure}[htbp]
\begin{center}
\includegraphics[width=1.0\linewidth]{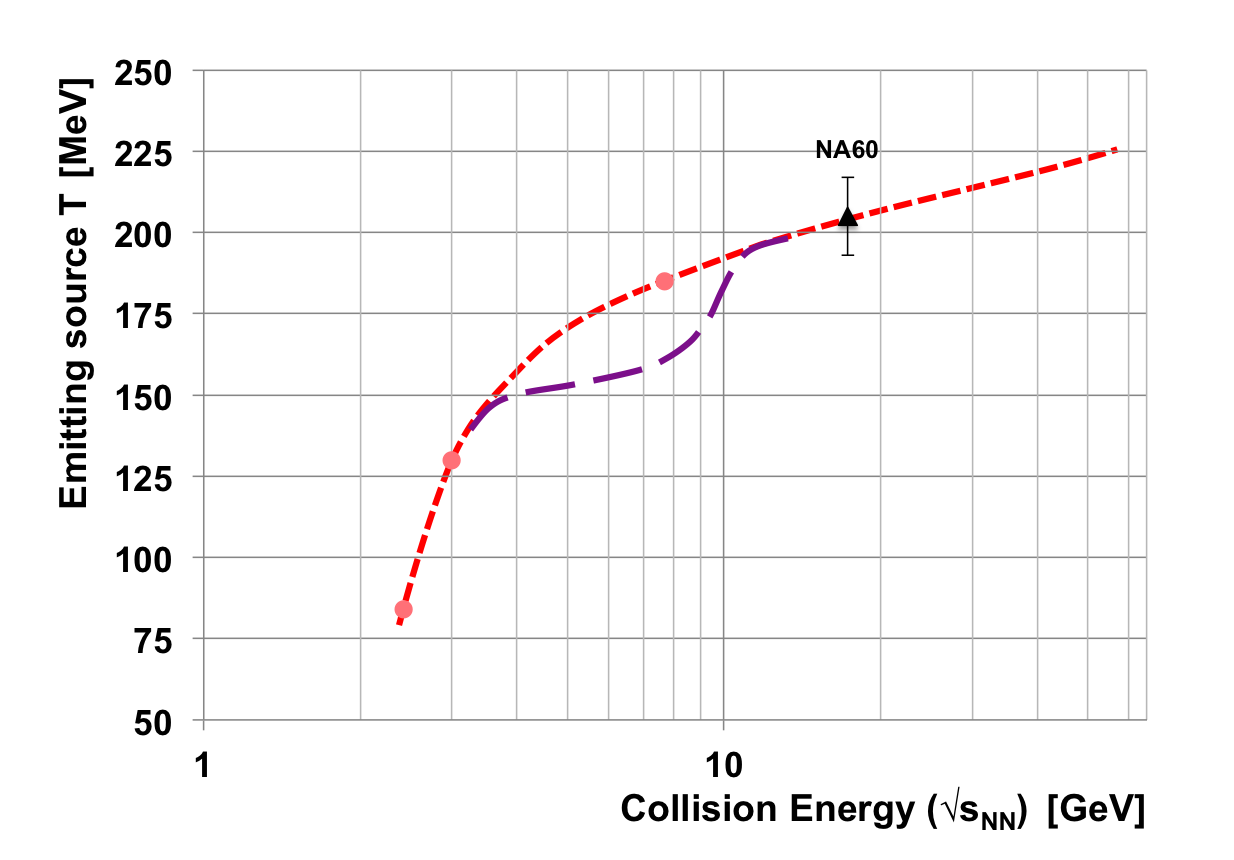}
\caption{
Excitation function of the fireball temperature $T$ extracted from intermediate dilepton mass distributions as calculated with a coarse-graining approach (dotted red curve)~\cite{seck15}. 
The dashed violet curve corresponds to a speculated shape with phase transition occurring in the SIS100 energy range. 
The black triangle corresponds to the temperature as measured by the NA60 collaboration at SPS~\cite{specht10}.
}
\label{fig:temperature}
\end{center}
\end{figure}

A very important part of the CBM research program will be the high-precision measurement of the di-lepton invariant mass distribution between 1 and 2.5~GeV/$c^2$ for different beam energies. 
With respect to top SPS, RHIC and LHC energies, the contribution of di-leptons from Drell-Yan processes or correlated charm decays, which also populate this mass region, are dramatically reduced at a beam energy of 20$A$ GeV as demonstrated in Fig.~\ref{fig:dielectrons} (note that at SIS100 energies these processes will contribute even less). 
This allows to access directly the fireball temperature and a contribution from $\rho$-a$_1$ chiral mixing. 
The precise measurement of the energy dependence of the spectral slope opens the unique possibility to measure the caloric curve, which would be the first direct experimental signature for phase coexistence in high-density nuclear matter. 
The excitation function of the fireball temperature $T$ extracted from the intermediate dilepton mass range, as calculated within the coarse-graining approach~\cite{seck15}, is shown in Fig.~\ref{fig:temperature} (red dotted curve). 
The dashed violet curve in Fig.~\ref{fig:temperature} shows a speculated shape of $T$ as function of collision energy, where the temperature saturates over a broad energy range. 
The flattening (plateau) of the caloric curve would clearly indicate a first-order phase transition, similar to the one presented as evidence for the liquid-gas phase transition in nuclear matter~\cite{agostino05}. 

In order to extract the continuum di-lepton signals, the physical and combinatorial background of lepton pairs has to be precisely determined, which is notoriously difficult. 
Since the background sources of electrons and muons are fundamentally different, independent measurements in both the di-electron and in the di-muon channel are decisive for the understanding of the systematic errors.

\subsection{Charm}

Particles containing charm quarks are expected to be created in the very first stage of the reaction, and, therefore, offer the possibility to probe the degrees-of-freedom over the entire collision history~\cite{averbeck13}. 
Depending on their interaction with the medium, the charm and anti-charm quarks hadronize into D mesons, charmed baryons, or charmonium. 
The suppression of charmonium due to colour screening of the heavy quark potential in the deconfined phase has been the first predicted signature for quark-gluon plasma formation~\cite{matsui86}. 
Charmonium suppression was first observed in central Pb+Pb collisions at 158$A$~GeV~\cite{abreu97}, and then also found in experiments at RHIC~\cite{adare07} and LHC~\cite{abelev12}. 
No data on open and hidden charm production in heavy-ion collisions are available at beam energies below 158$A$~GeV. Moreover, the interpretation of existing data is complicated by lacking knowledge of interactions between charmed particles and the cold hadronic medium~\cite{kharzeev95}.

With CBM at SIS100, charm production will be studied for the first time at beam energies close to production threshold. 
At these energies, the formation time of charmonium is small compared to the lifetime of the reaction system. 
CBM is thus uniquely suited to study the interactions between fully formed J/$\psi$ and the dense medium with appropriate counting statistics and systematics. 

Systematic measurements of charmonium in p+A collisions with varying target mass number A at proton energies up to 30~GeV will shed light on the charmonium interaction with cold nuclear matter and constitute an important baseline for measurements in nuclear collisions. 
Moreover, the simultaneous measurement of open charm will give access to the basically unknown charm production cross section at or near the kinematic threshold. 
Based on simulations with the HSD event generator~\cite{cassing01}, the yield of D mesons and charmonium expected in p+A collisions at SIS100 energies after a run of 10 weeks varies between $10^4$ and $10^6$, depending on proton energy, and is sufficient to perform a multi-differential analysis.

CBM will also extend the measurement of the J/$\psi$ as probe of the hot medium to the lower energies. 
At SIS100, charmonium will be measured in collisions of symmetric nuclei up to 15$A$~GeV and, more challenging even, below threshold in Au+Au collisions at 10$A$~GeV. 
Model predictions of the J/$\psi$ multiplicity in this energy range vary widely. 
Taking the prediction of the HSD model~\cite{cassing01}, the yield obtained in one week of running at an interaction rate of 10~MHz would be about 300 J/$\psi$ for central Au+Au collisions at 10$A$~GeV, and about 600 J/$\psi$ for central Ni+Ni collisions at 15$A$~GeV. 
In the latter case, also open charm production can be studied. 
However, because of the rate limitations of the MVD which is needed to select the D meson decay vertex, the measurement will be performed at a rate of 300~kHz. 
As a result, the expected yield in central Ni+Ni collisions at 15$A$~GeV will be about 30 D mesons per week. 
This would be sufficient for an analysis of charmonium propagation and absorption in dense baryonic matter based on the ratio of hidden to open charm.

\subsection{Hypernuclei and strange objects}

\begin{figure}[htbp]
\begin{center}
\includegraphics[width=1.0\linewidth]{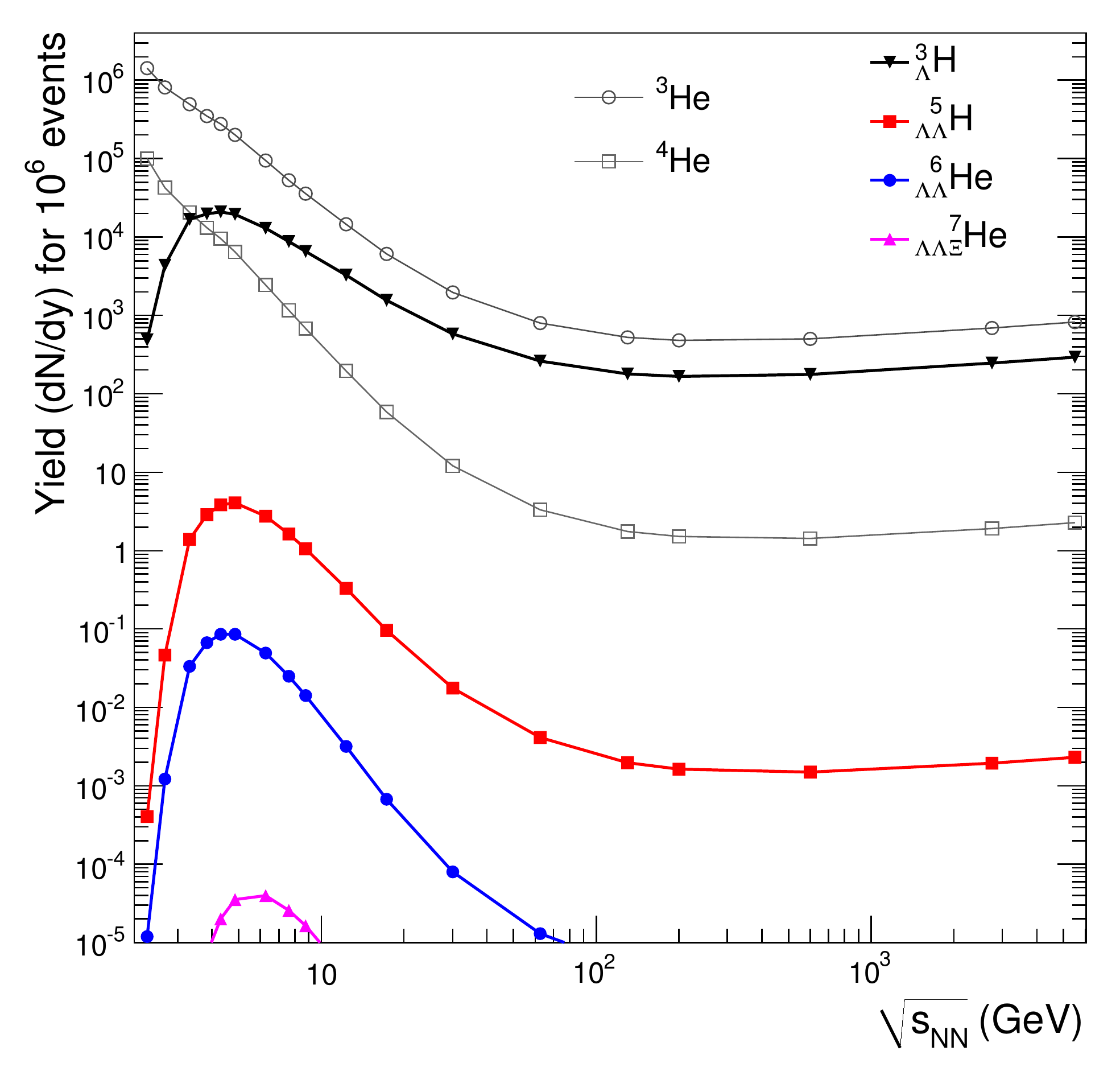}
\caption{
Energy dependence of hypernuclei yields at midrapidity for $10^6$ central collisions as calculated with a thermal model. 
The predicted yields of $^3$He and $^4$He nuclei are included for comparison~\cite{andronic11}.
}
\label{fig:hypernuclei}
\end{center}
\end{figure}

Thermal model calculations predict the production of single and double hypernuclei in heavy-ion collisions~\cite{andronic11}.  
The results of these calculations are shown in Fig.~\ref{fig:hypernuclei} demonstrating that the excitation function of hypernucleus production exhibits its maximum in the SIS100 energy range. 
This is due to the superposition of two effects: the increase of light nuclei production with decreasing beam energy, and the increase of hyperon production with increasing beam energy.

The CBM experiment at SIS100 will measure hydrogen and helium hypernuclei in huge amounts. 
Moreover, the experiment has a substantial discovery potential for light double-$\Lambda$ hypernuclei. 
According to Fig.~\ref{fig:hypernuclei}, in 1 million central Au+Au collisions the hypernuclei $\prescript{5}{\Lambda\Lambda}{\mathrm{H}}$ and $\prescript{6}{\Lambda\Lambda}{\mathrm{He}}$ will be produced at a beam energy around 10$A$~GeV with a yield of about 5 and 0.1, respectively. 
Assuming a reaction rate of $10^6$ central events/s, a branching ratio of 10\% for two sequential weak decays, and an efficiency of 1\%, one would expect to measure within one week about  3000 $\prescript{5}{\Lambda\Lambda}{\mathrm{H}}$ and 60 $\prescript{6}{\Lambda\Lambda}{\mathrm{He}}$, respectively.  
Such measurements would represent a breakthrough in hypernucleus physics, as up to now only very few double-$\Lambda$ hypernuclei events have been found~\cite{ahn13}.
 The discovery of (double-) $\Lambda$ hypernuclei and the determination of their lifetimes will provide information on the hyperon-nucleon and hyperon-hyperon interactions, which are essential ingredients for the understanding of the nuclear matter equation-of-state at high densities, and, hence, of the structure of neutron stars~\cite{botvina14}.

According to a coupled transport-hydro-dynamics model, the high baryon densities created in heavy-ion collisions at FAIR energies favour the distillation of strangeness~\cite{stoecker09}. 
The model predicts the production of hypernuclei, strange di-baryons, and multi-strange short-lived objects. 
These predictions open the exciting perspective to explore the formation of composite objects with multiple strangeness in heavy-ion collisions at SIS100 energies.

\section{Experiments with CBM detectors in ``FAIR Phase 0''}
\label{sec:phase0}

The start version of the CBM experiment will be ready to take the first beam from SIS100 in the year 2024. 
However, several detector and software components will be ready earlier. 
Therefore, it is planned to install some of these components in existing heavy-ion experiments at other laboratories. 
The benefits of these efforts are manifold: The additional detectors will improve the performance of the experiments, the CBM detectors and their readout chain will be commissioned and calibrated, 
the reconstruction and analysis software will be tested and advanced on real data,
and members of the CBM collaboration will participate in data taking and analysis, thereby maintaining experience in performing physics measurements and educating the next generation of young physicists. 
The projects are briefly sketched in the following. 

The photon detector of the RICH detector in HADES will be replaced by modern Multi-Anode Photo-Multipliers (MAPM) which have been ordered for the CBM RICH detector. 
The CBM RICH detector comprises 1100 MAPMs out of which 428 will be installed in HADES. 
The new detector will substantially improve in particular the di-lepton pair efficiency for small opening angles, and, hence, the electron identification performance of the HADES experiment for the intermediate research program at GSI between 2018 and the start of FAIR. 
After commissioning of CBM, a part of the MAPMs will be shared by both experiments applying an alternating running scenario.

About 10\% of the TOF-MRPC modules will be installed at the STAR detector at RHIC. 
They will serve as an endcap TOF-wall in order to increase the particle identification capability at forward rapidities. 
According to current planning, 36 CBM TOF modules housing 108 MRPC detectors will cover an active area of about 10 m$^2$ and comprise 10.000 read-out channels. 
An installation of a prototype module is planned already for fall 2016 and is necessary in order to develop the interface of the different data acquisition systems of CBM and STAR. 
The installation of the full set of MRPC detector modules is planned to start in spring 2018 allowing the participation in the Beam-Energy Scan II at RHIC in 2019/20, where STAR will run both in the collider and the fixed target mode.
Although the interaction rates will be relatively low, the CBM system will be exposed to particle multiplicities close to the ones expected for running at SIS100. 
For STAR the extension of acceptance is vital to improve the particle identification coverage for a number of interesting bulk observables like the rapidity density of antiprotons, the directed and elliptic flow of protons and pions, and the measurement of event-by-event fluctuations of net-baryons. 
Also for the strangeness program targeting among others on the $v_2$ measurement of $\phi$ mesons, visible contributions of the CBM-TOF subsystem are anticipated. 
CBM members will gain operational experience in running the subsystem and will be involved in the data analysis and physics publications.

The CBM track finding algorithm based on the Cellular Automaton will be used in data production as well as in the High-Level Trigger (HLT) of the STAR experiment. 
Similarly, the KF Particle package for the analysis of short-lived particles will be used for online event selection and offline physics analysis in STAR. 
These new algorithms will improve the track reconstruction efficiency and the data processing speed,
and will enable online physics analysis on the STAR HLT computer farm equipped with many-core CPUs and accelerator cards. 
The reconstruction algorithms will be used by the \mbox{ALICE} experiment at CERN as well.
CBM members will thus take part in data taking, physics analysis and in publications of these experiments.

Four prototype stations of the CBM Silicon Tracking System (STS) are considered to be installed in the future fixed-target  experiment BM@N at the Nuclotron at JINR in Dubna. 
The construction of the stations is planned as a joint venture of the CBM-STS group and the BM@N collaboration. 
The silicon stations will significantly increase the track reconstruction efficiency in particular for low particle momenta, and, therefore, improve the performance for the identification of multi-strange hyperons, which are the most important observables of the physics program of BM@N. 
The participation in this experiment will be very valuable in commissioning the CBM Silicon detector itself and for the development of tracking and physics analysis strategies under experimental conditions. 
Data taking with Au-beams of energies up to 4.5$A$~GeV at moderate rates is planned from 2018 -- 2021. 

A number of tests of the CBM Projectile Spectator Detector (PSD) components are planned and/or currently ongoing at the NA61/SHINE facility. 
The readout electronics, the response of the hadron calorimeter modules, and the PSD performance for collision geometry determination (centrality and event plane) are being under investigation with a similarly designed PSD of NA61/SHINE.

At GSI/SIS18 we plan to install a setup consisting of full-size CBM detector modules including the data readout chain up to the GreenIT cube to perform system tests with high-rate nucleus-nucleus collisions. 
These measurements will allow to optimize the performance of the detectors under experiment conditions and to test the free-streaming data transport including the online event selection on a high-performance computing cluster. 
The goal is to reduce the time for CBM commissioning at SIS100. 
The setup will be installed in 2017/2018.

\section{Conclusions}
\label{sec:conclusions}

In heavy-ion collisions at beam energies available at SIS100, model calculations predict the creation of strongly interacting QCD matter at extreme values of density, similar to neutron star core densities. 
This offers the opportunity to explore the QCD phase diagram in the region of high net-baryon densities, to study the equation of state, to search for phase transitions, chiral symmetry restoration, and exotic forms of (strange) QCD matter with a dedicated experiment.

The CBM detector is designed to measure the collective behaviour of hadrons, together with rare diagnostic probes such as multi-strange hyperons, charmed particles and vector mesons decaying into lepton pairs with unprecedented precision and statistics. 
Most of these particles will be studied for the first time in the FAIR energy range. 
In order to achieve the required precision, the measurements will be performed at reaction rates up to 10~MHz. 
This requires very fast and radiation hard detectors, a novel data read-out and analysis concept including free streaming front-end electronics, and a high performance computing cluster for online event selection.
Several of the CBM detector systems, the data read-out chain and event reconstruction will be commissioned and already used in experiments  during the FAIR phase 0.

The unique combination of an accelerator which delivers a high-intensity heavy-ion beam with a modern high-rate experiment based on innovative detector and computer technology offers optimal conditions for a research program with substantial discovery potential for fundamental properties of QCD matter.


\end{document}